\documentclass[preprint,12pt,sort&compress]{elsarticle}
\usepackage{graphicx}
\usepackage{amsmath}
\begin{document}
\newcommand {\nc} {\newcommand}
\nc {\beq} {\begin{eqnarray}}
\nc {\eol} {\nonumber \\}
\nc {\eeq} {\end{eqnarray}}
\nc {\eeqn} [1] {\label{#1} \end{eqnarray}}
\nc {\eoln} [1] {\label{#1} \\}
\nc {\ve} [1] {\mbox{\boldmath $#1$}}
\nc {\rref} [1] {(\ref{#1})}
\nc {\Eq} [1] {Eq.~(\ref{#1})}
\nc {\re} [1] {Ref.~\cite{#1}}
\nc {\dem} {\mbox{$\frac{1}{2}$}}
\nc {\arrow} [2] {\mbox{$\mathop{\rightarrow}\limits_{#1 \rightarrow #2}$}}

\author[ulb]{D. Baye\corref{cor}}
\ead{dbaye@ulb.ac.be}
\author[inp]{E.M. Tursunov}
\ead{tursune@inp.uz}
\address[ulb]{Physique Quantique, CP 165/82, and Physique Nucl\'eaire Th\'eorique et Physique Math\'ematique, CP229,
Universit\'e Libre de Bruxelles (ULB), B-1050 Brussels, Belgium}
\address[inp]{Institute of Nuclear Physics, Uzbekistan Academy of Sciences, \\ 
100214, Ulugbek, Tashkent, Uzbekistan}
\cortext[cor]{Corresponding author}
\title{$\beta$ delayed emission of a proton by a one-neutron halo nucleus}
\date{\today}
\begin{abstract}
Some one-neutron halo nuclei can emit a proton in a $\beta$ decay of the halo neutron. 
The branching ratio towards this rare decay mode is calculated within a two-body potential model 
of the initial core+neutron bound state and final core+proton scattering states. 
The decay probability per second is evaluated for the $^{11}$Be, $^{19}$C and $^{31}$Ne one-neutron halo nuclei. 
It is very sensitive to the neutron separation energy. 
\end{abstract}
\maketitle
\section{Introduction}
Some neutron-rich halo nuclei can emit a proton. 
This process is possible if the neutron separation energy is very small. 
Indeed a weakly bound halo neutron may $\beta$ decay, 
producing a proton which can be emitted, possibly together with neutrons. 
Processes where this proton is bound with one or two neutrons have been observed 
in the $\beta$ delayed deuteron and triton decays of $^6$He and $^{11}$Li 
\cite{LDE84,RBG90,BJJ93,ABB02,MBG96,BGG97,RAB08,MBA09}. 
Recently we have calculated the branching ratio of an even rarer process 
where the proton remains unbound but is accompanied by a free neutron \cite{BDT10}. 
This decay is uniquely possible for $^{11}$Li, among nuclei with known separation energies. 
The study has been performed in a three-body model with a simplified description of the continuum. 
An even simpler process is however possible. 

A one-neutron halo nucleus can be viewed as a normal nucleus, the core, 
to which a neutron is bound in an orbital with a large radius. 
The $\beta$ decay of the bound halo neutron may occur, releasing the proton, 
under the condition of energy conservation 
\beq
S_{\rm n} < (m_n - m_p -m_e) c^2 \approx 0.782 {\rm\ MeV},
\eeqn{1.1}
where $S_{\rm n}$ is the neutron separation energy of the decaying nucleus 
and $m_n$, $m_p$ and $m_e$ are the neutron, proton and electron masses, respectively. 
Among one-neutron halo nuclei for which $S_n$ is known with sufficient precision, 
this decay is allowed at least for $^{11}$Be and $^{19}$C, and probably for $^{31}$Na.  
It should be observable if the branching ratio is large enough. 
This decay mode of $^{11}$Be has been considered by Horoi and Zelevinsky 
but the results do not seem to have been published \cite{HZ03}. 
Here we study this rare decay mode within a two-body potential model. 
The initial halo nucleus is treated as a core+neutron bound state. 
The final states lie in the core+proton continuum. 
How rare is this decay is the main question raised in the present exploratory study. 
\section{Decay probability for $\beta$ delayed proton emission}
\label{sec:theory}
The $\beta$ decay of the halo neutron releases the resulting proton from the core. 
The distribution of decay probability per time unit 
as a function of the energy $E < Q$ of the relative motion 
of the two particles is given by 
\beq
\frac{dW}{dE} = \frac{1}{2\pi^3} \frac{m_e c^2}{\hbar} G_{\beta}^2 f(Q-E) 
\left( \frac{dB(\rm F)}{dE} + \lambda^2 \frac{dB(\rm GT)}{dE} \right),
\eeqn{2.1}
where $G_{\beta} \approx 2.996 \times 10^{-12}$ is the dimensionless $\beta$-decay constant and 
$\lambda \approx -1.268$ is the ratio of the axial-vector to vector coupling constants. 
The Fermi integral $f(Q-E)$ depends on the kinetic energy $Q-E$ available for the electron and antineutrino 
with 
\beq
Q = (m_n - m_p -m_e) c^2 - S_{\rm n}.
\eeqn{2.2}
The total decay probability per time unit $W$ is obtained by integrating \rref{2.1} from zero to $Q$. 
The branching ratio can than be derived as 
\beq
{\cal R} = W t_{1/2}/\ln 2,
\eeqn{2.3}
where $t_{1/2}$ is the half life of the halo nucleus. 

In the present model, the halo nucleus is described as a two-body core+ neutron 
system in its ground state with total angular momentum $J_i$ 
resulting from the coupling of the orbital momentum $l_i$ of the relative motion 
and the neutron spin $s=1/2$. 
The spin of the core is assumed to be zero. 
The parity of the initial state is $(-1)^{l_i}$. 
The radial wave function is denoted as $u_{i l_i J_i}$ with the normalization 
$\int_0^{\infty} |u_{i l_i J_i}(r)|^2 dr = 1$. 
It is obtained from a potential $V_i$ adjusted to reproduce the experimental 
neutron separation energy $S_{\rm n}$. 

The final scattering state of the core and the proton is a distorted wave 
with wave vector $\ve{k}$. 
Because of selection rules, only some partial waves with total angular momentum $J_f$ 
resulting from the coupling of the orbital momentum $l_f$ and the proton spin $s$ are allowed. 
The radial wave functions $u_{k l_f J_f}$ for a wave number $k = \sqrt{2\mu E/\hbar^2}$ 
where $\mu$ is the core-proton reduced mass 
are obtained with a potential $V_f$ describing the core+proton system. 
They are normalized according to 
$\int_0^{\infty} u_{k l_f J_f}(r) u_{k' l_f J_f}(r) dr = \delta(k-k')$. 
The potential $V_f$ is usually poorly known when the core is unstable. 

Within this model, the Fermi reduced decay probability is given by 
\beq
\frac{dB({\rm F})}{dE} = \frac{1}{\hbar v} 
\left| I_{l_i J_i J_i} \right|^2
\eeqn{4.1}
and the Gamow-Teller reduced decay probability by 
\beq
\frac{dB({\rm GT})}{dE} = \frac{6}{\hbar v} 
\sum_{J_f} (2J_f+1) \left\{ \begin{array}{ccc}
J_f &  s  & l_i \\
s   & J_i & 1
\end{array} \right\}^2 
\left| I_{l_i J_i J_f} \right|^2
\eeqn{4.2}
with the relative velocity $v = \hbar k/\mu$ and the radial integrals 
\beq
I_{l J_i J_f} = \int_0^\infty u_{k l J_f} (r) u_{i l J_i} (r) dr.
\eeqn{4.3}

If the final wave function does not depend on $J_f$, 
the Gamow-Teller term simplifies as 
\beq
\frac{dB({\rm GT})}{dE} = 3 \frac{dB({\rm F})}{dE}.
\eeqn{4.4}
The reduced decay probability can then also be written as 
\beq
\frac{dW}{dE} = W_n\, \frac{f(Q-E)}{f_n}\, \frac{dB(\rm F)}{dE},
\eeqn{4.5}
where $W_n$ is the free-neutron $\beta$ decay probability per second 
and $f_n$ is the corresponding Fermi integral. 

With respect to a free neutron, the decay probability is affected in two ways. 
First, the ratio $f(Q-E)/f_n$ is small due to the reduction of phase space, 
since $f_n \equiv f(Q + S_{\rm n})$. 
It becomes extremely small when $E$ tends to $Q$. 
The $\beta$ delayed proton emission is favoured by very small separation energies $S_{\rm n}$. 
Second, the reduced decay probability \rref{4.1} appearing in \rref{4.5} 
is proportional to the square of a radial integral \rref{4.3}. 
Because of the Coulomb repulsion and the smallness of the $Q$ value, 
the scattering waves are small and, when $E$ tends to zero, tend to zero 
as $k^{1/2} \exp(-\pi\eta)$  \cite{BB00}, 
where $\eta = Z_c e^2/\hbar v$ is the Sommerfeld parameter. 
They become thus smaller with increasing charge $Z_c$ of the core. 
They also become smaller with increasing orbital momentum. 
Hence, at given $Q$ value, we expect the decay probability to be 
largest for the lightest halo nuclei and for the halo neutron in the $s$ wave. 
\section{Results and discussion}
\label{sec:results}
Before making explicit calculations, we have to specify the choice of potentials. 
The Fermi strength is proportional to the square of an overlap integral \rref{4.3} 
between the initial and final radial wave functions. 
In order to have a realistic overlap, 
it is useful to have a correct node structure for these wave functions. 
Indeed, the presence of nodes leads to an integrand that changes 
sign one or several times and thus to a reduction of the overlap. 
Spectroscopic factors can also affect the size of the Fermi strength 
but given the limited knowledge on these quantities, we choose 
to ignore them in the present exploratory study. 
Finally, absorption in the core+proton optical potential might also 
play a role. 
However, the energies of the states after decay are lower than, or comparable to, 
the energy of the Coulomb barrier. 
Absorption should be weak and can safely be neglected. 

Hence, we shall use real potentials $V_i$ and $V_f$ 
which should be deep enough to provide a realistic node structure 
of the initial and final radial wave functions. 
To keep the model simple we only use central Woods-Saxon potentials 
with range $r_0 A_c^{1/3}$ where $A_c$ is the mass number of the core. 
The depth is adapted to the separation energy for the core+n system. 
The same form factor with an additional point-sphere Coulomb potential 
is employed for the final core+p elastic scattering. 
Because of the small energies, the phase shifts are small and 
the sensitivity to $V_f$ is weak. 
Now let us consider explicit cases. 

The best documented case is $^{11}$Be. 
Its $1/2^+$ ground state has a separation energy of about 501 keV \cite{RBB09} 
and its half life is 13.8 s \cite{ABB03}. 
The halo neutron is described by an $s$ wave. 
The parameters of the Woods-Saxon potential are taken 
as $r_0 = 1.2$ fm, $a = 0.6$ fm and $V_{i0} = 62.52$ MeV \cite{CGB04}. 
In the $s$ wave, this potential possesses one unphysical forbidden state. 
The same parameters are used for the final potential except $V_{f0}$. 
The $^{11}$B nucleus has a proton separation energy $S_{\rm p} \approx 11.228$ MeV \cite{AWT03}. 
Its lowest $1/2^+$ state is located at the excitation energy $E_x \approx 6.79$ MeV. 
In the $s$ wave, $V_{f0} = 84.1$ MeV is adjusted so that the potential possesses 
one forbidden state and one bound state fitted to the energy $E_x - S_{\rm p} \approx -4.52$ MeV 
with respect to the $^{10}$Be+p threshold. 
Bound and scattering states should thus have a reasonable node structure. 
\begin{figure}[ht]
\setlength{\unitlength}{1mm}
\begin{picture}(140,85) (0,0) 
\put( -5,-60){\mbox{\scalebox{0.5}{\includegraphics{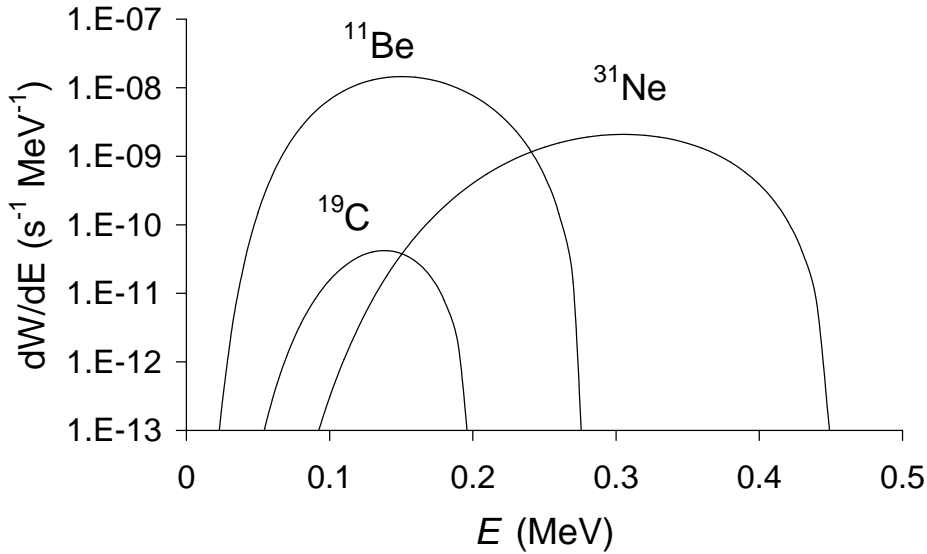}}}}
\end{picture} \\
\caption{Distribution of decay probability per second 
for the $\beta$ delayed $np$ decay of $^{11}$Be, $^{19}$C and $^{31}$Ne.}
\label{fig:1}
\end{figure}

The $Q$ value \rref{2.2} is small, 0.281 MeV. 
The distribution of decay probability is displayed in Fig.~\ref{fig:1}. 
The most probable energies of the relative motion are in the interval 0.1-0.2 MeV. 
The total decay probability $1.5 \times 10^{-9}$ s$^{-1}$ leads 
to a branching ratio $3.0 \times 10^{-8}$. 

The $^{19}$C $1/2^+$ ground state has a separation energy of $580 \pm 90$ keV \cite{AWT03} 
and a half life $t_{1/2} = 46.2$ ms \cite{ABB03}. 
As a simple picture, we consider a neutron in the $s$ wave with one forbidden state 
and no spectroscopic factor.  
The parameters of the Woods-Saxon potential are 
$r_0 = 1.25$ fm, $a = 0.62$ fm and $V_{i0} = 41.42$ MeV
giving a $Q$ value of 202 keV. 
For the final $^{18}$C+p system, the $s$ wave possesses one forbidden state. 
We assume a possible $1/2^+$ bound state near $E_x = 2.1$ MeV \cite{SSD08}. 
With $S_{\rm p} \approx 16.35$ MeV \cite{AWT03}, 
we take $V_{f0} = 77.2$ MeV which gives a bound state at $-14.2$ MeV. 
The distribution of decay probability is displayed in Fig.~\ref{fig:1}. 
It is much smaller than for $^{11}$Be because of the larger charge of the 
core and the smaller $Q$ value. 
The total decay probability $2.7 \times 10^{-12}$ s$^{-1}$ leads 
to a branching ratio $1.8 \times 10^{-13}$. 

A candidate for delayed proton emission is $^{31}$Ne. 
Its neutron separation energy is poorly known: $0.33 \pm 1.07$ MeV \cite{AWT03}. 
Its half life is $t_{1/2} = 3.4$ ms \cite{ABB03}. 
This nucleus belongs to an island of inversion where its ground state should be an intruder state. 
Its one-neutron removal cross section \cite{NKK09} is too large for agreeing with the quantum 
numbers $0f7/2$ of the naive shell model. 
The ground state could be described with a $1p3/2$ orbital \cite{HSC10} 
although a $2s1/2$ orbital has also been considered \cite{Ha10}. 
Here we assume a $p$ wave ground state at $-0.33$ MeV giving $Q = 0.45$ MeV. 
It can be reproduced with the parameters $r_0 = 1.25$ fm, $a = 0.75$ fm 
and $V_{i0} = 48.86$ MeV \cite{HSC10}. 
This potential has one forbidden state in the $p$ wave. 
Little is known about the $^{30}$Ne+p scattering. 
One can also expect an intruder $3/2^-$ state in the vicinity of the ground state. 
Hence we choose $V_{f0} = 90.0$ MeV which provides a forbidden state 
and a bound state at $-16.1$ MeV, not far above $-S_{\rm p} \approx -17.7$ MeV. 

The distribution of decay probability is displayed in Fig.~\ref{fig:1}. 
It is smaller than for $^{11}$Be because of the larger charge of the core and 
the higher orbital momentum, but these effects are partly compensated by the larger $Q$ value. 
The most probable energies $E$ lie between 0.25 and 0.35 MeV. 
The total decay probability $3.3 \times 10^{-10}$ s$^{-1}$ leads 
to a branching ratio $1.6 \times 10^{-12}$. 
For an $s$ ground state with two forbidden states ($V_{i0} = 69.27$ MeV), 
the decay probability $W \approx 1.6 \times 10^{-9}$ would be five times larger. 

\begin{figure}[t]
\setlength{\unitlength}{1mm}
\begin{picture}(140,70) (0,0) 
\put(-10,-60){\mbox{\scalebox{0.5}{\includegraphics{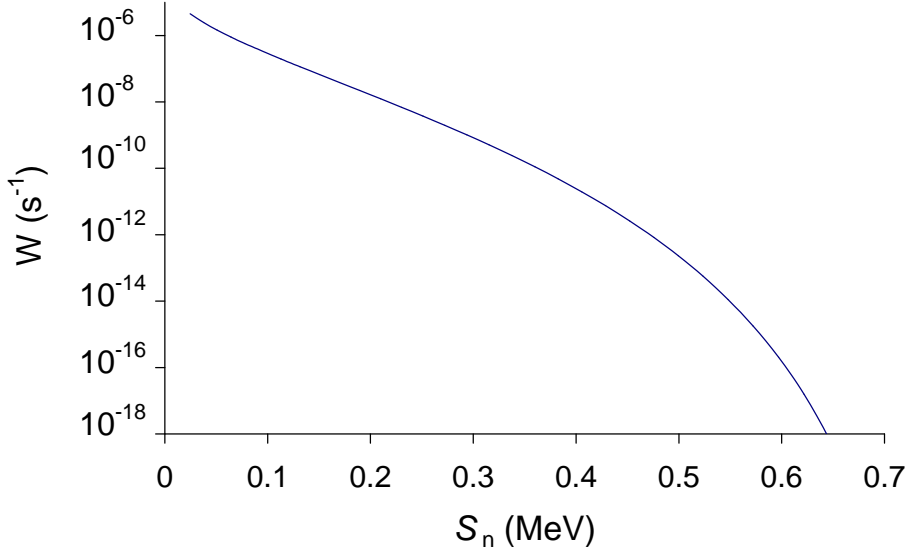}}}}
\end{picture} \\
\caption{Decay probability per second 
for the $\beta$ delayed $np$ decay of $^{31}$Ne 
as a function of the separation energy $S_{\rm n}$.}
\label{fig:2}
\end{figure}
The separation energy of $^{31}$Ne is quite uncertain. 
The one-neutron removal cross section can be interpreted as arising from $S_{\rm n} \approx 0.6$ MeV 
but this assumption is weakened by the lack of knowledge of spectroscopic factors \cite{HSC10}. 
Hence we display in Fig.~\ref{fig:2} the dependence of the decay probability 
on the separation energy $S_{\rm n}$. 
One observes that it varies very strongly. 
If $S_{\rm n}$ is around 0.6 MeV, the decay probability is reduced 
by about six orders of magnitude. 
On the contrary, the decay probability can be larger by five orders of magnitude 
if the separation energy is very small. 

Finally, let us note that an estimate of the order of magnitude 
(in general within a factor of two) can be obtained with the 
simple approximation 
\beq
I_{l J_i J_f} = C \sqrt{\frac{2}{\pi}}\, \int_a^\infty F_l (\eta,kr) e^{-\kappa r} dr
\eeqn{4.6}
where $a = 5$ fm, $\kappa = \sqrt{2\mu S_n/\hbar^2}$ and $F_l$ is a regular Coulomb function. 
Under the same conditions as in Fig.~\ref{fig:1}, 
the asymptotic normalization constant $C$ is 0.83, 0.96, 0.69 fm$^{-1/2}$ 
for $^{11}$Be, $^{19}$C, $^{31}$Ne, respectively. 
\section{Conclusion}
\label{sec:conclusion}
As a summary, we have evaluated the order of magnitude of the decay probability 
per second for the $\beta$ delayed proton emission by one-neutron halo nuclei. 
The best candidate for observing such a decay is $^{11}$Be 
in spite of the fact that its separation energy is not very small. 
The probability of this delayed decay is smaller than for the neutron-and-proton delayed 
decay of $^{11}$Li \cite{BDT10} by an order of magnitude. 
Because of a longer lifetime, the branching ratio is larger by two orders of magnitude. 
The observation of this $\beta$ delayed decay mode of $^{11}$Be 
would thus require high radioactive beam intensities and long measurement times. 

The neutron separation energies of the other candidates, $^{19}$C and $^{31}$Ne, 
are less well known and the decay probabilities are thus more uncertain. 
We have shown that the decay probability varies strongly with the neutron separation energy. 
A very small $S_{\rm n}$ would be advantageous for the study of this decay mode. 
This advantage however decreases when the charge of the core increases. 
The best candidate for observing such a decay would be a not too heavy one-neutron halo nucleus 
with a very small separation energy. 
\section*{Acknowledgments}
This text presents research results of BriX (Belgian Research Initiative on eXotic nuclei), 
the interuniversity attraction pole programme P6/23 initiated by the Belgian-state 
Federal Services for Scientific, Technical and Cultural Affairs (FSTC). 
E.M.T. thanks the IAP programme for supporting his stay.
%
%

%
\end{document}